\documentclass[10pt,twocolumn,letterpaper]{article}

\usepackage[pagenumbers]{cvpr} % To force page numbers, e.g. for an arXiv version
\usepackage{amsmath}
\usepackage{algorithm}
\usepackage{algorithmic}
\usepackage{color} 
\usepackage{listings}
\usepackage[dvipsnames]{xcolor}
\usepackage[accsupp]{axessibility} % Improves PDF readability for those with visual impairments.

\usepackage[dvipsnames]{xcolor}

\definecolor{cvprblue}{rgb}{0.21,0.49,0.74}
\usepackage[pagebackref,breaklinks,colorlinks,citecolor=cvprblue]{hyperref}
\usepackage{multirow}
\def\paperID{7530} % *** Enter the Paper ID here
\def\confName{CVPR}
\def\confYear{2024}

\makeatletter
\def\blfootnote{\gdef\@thefnmark{}\@footnotetext}
\makeatother
\title{MicroDiffusion: Implicit Representation-Guided Diffusion for 3D Reconstruction from Limited 2D Microscopy Projections}

\author{Mude Hui\textsuperscript{\rm 1}\thanks{Denotes equal first author} \quad  Zihao Wei\textsuperscript{{\rm 2}}\footnotemark[1]  \quad  Hongru Zhu\textsuperscript{{\rm 3}} \quad Fei Xia\textsuperscript{{\rm 4}}\thanks{Denotes equal senior author} \quad Yuyin Zhou\textsuperscript{\rm 1}\footnotemark[2]\\
\textsuperscript{\rm 1}{University of California, Santa Cruz}\quad
\textsuperscript{\rm 2}{University of Michigan, Ann Arbor}\quad\\
\textsuperscript{\rm 3}{Johns Hopkins University} \quad
\textsuperscript{\rm 4}{Ecole Normale Supérieure de Paris} \\
\tt\small muhui@ucsc.edu
\tt \small  zihaowei@umich.edu
\tt \small \{hongruzhu95,\ zhouyuyiner\}@gmail.com
\tt \small fx43@cornell.edu}

\begin{document}

\maketitle
\begin{abstract}
% To address these limitations in imaging hardware and existing INR techniques, here we present MicroDiffusion, a technique that refines INR outputs by incorporating the Denoising Diffusion Probabilistic Model (DDPM), known for preserving fine image details by leveraging a U-Net model. 
% MicroDiffusion uses INR as a prior to guide the diffusion process.
% It introduces a novel initialization strategy by blending the INR model's output with a perturbed input via linear interpolation, rather than starting from the conventional Gaussian noise baseline. This initialization enriches the diffusion process with 3D structured information.
% MicroDiffusion utilizes a cross-attention mechanism that dynamically integrates information from the 2D snapshot with features extracted by a U-Net architecture. This process allows for a highly informed and specific refinement of the 3D reconstruction, ensuring fidelity to the original image details.
% the use of INR as a global prior for structural coherence, coupled with DDPM's prowess in fine detail enhancement. 
Volumetric optical microscopy using non-diffracting beams enables rapid imaging of 3D volumes by projecting them axially to 2D images but lacks crucial depth information. Addressing this, we introduce MicroDiffusion, a pioneering tool facilitating high-quality, depth-resolved 3D volume reconstruction from limited 2D projections. 
While existing Implicit Neural Representation (INR) models often yield incomplete outputs and Denoising Diffusion Probabilistic Models (DDPM) excel at capturing details, our method integrates INR's structural coherence with DDPM's fine-detail enhancement capabilities. We pretrain an INR model to transform 2D axially-projected images into a preliminary 3D volume. This pretrained INR acts as a global prior guiding DDPM's generative process through a linear interpolation between INR outputs and noise inputs. This strategy enriches the diffusion process with structured 3D information, enhancing detail and reducing noise in localized 2D images.
By conditioning the diffusion model on the closest 2D projection, MicroDiffusion substantially enhances fidelity in resulting 3D reconstructions, surpassing INR and standard DDPM outputs with unparalleled image quality and structural fidelity. Our code and dataset are available at~\url{https://github.com/UCSC-VLAA/MicroDiffusion}.

\end{abstract}    
\section{Introduction}
\begin{figure}[htbp]
\centering
\includegraphics[width=1.0\linewidth]{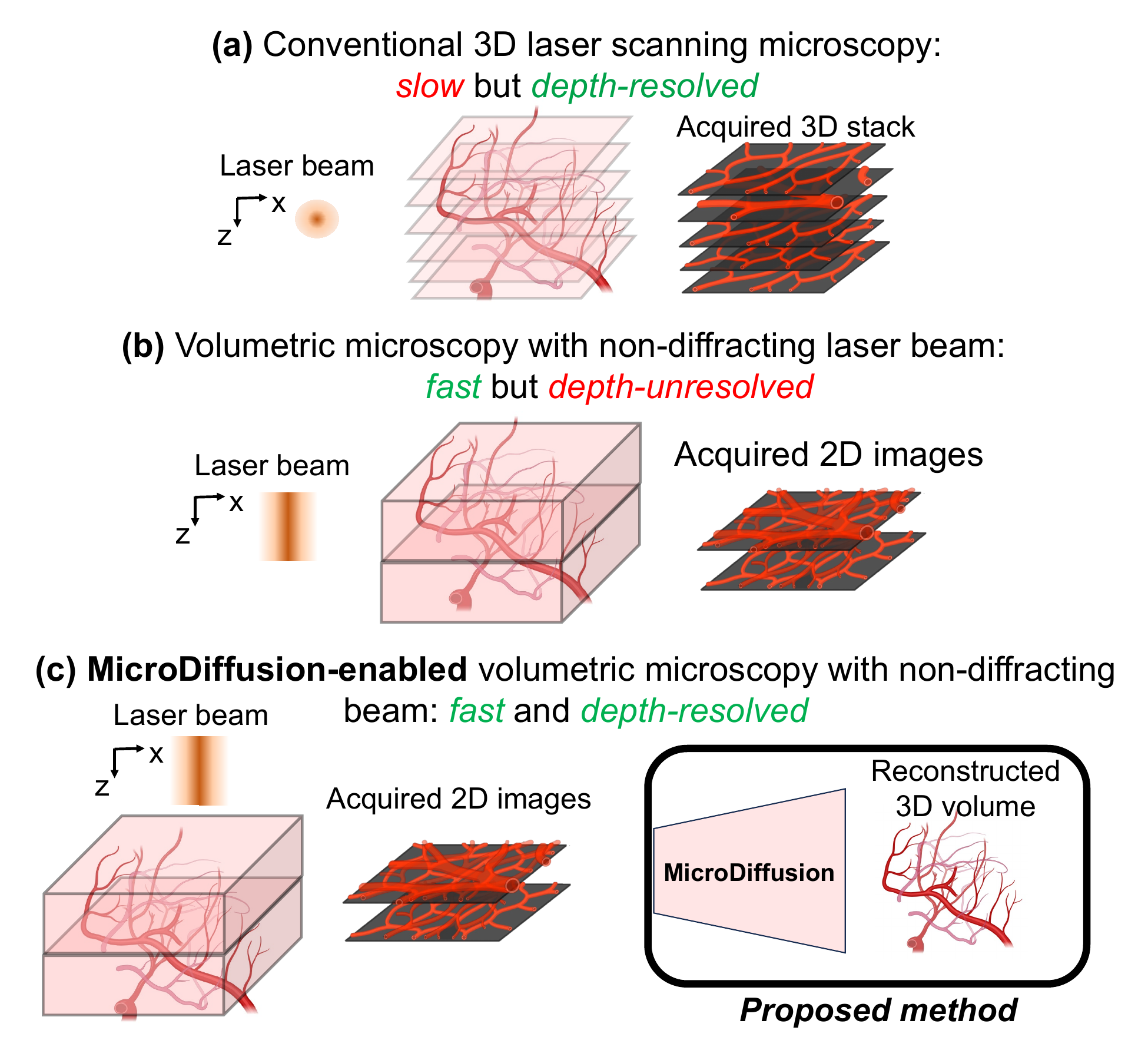}
%\caption{The MicroDiffusion-enabled volumetric microscopy concept addresses limitations in traditional 3D laser scanning and volumetric microscopy. (a) Conventional 3D laser scanning, though depth-resolvable, is slow for volumetric imaging. (b) Volumetric microscopy uses a non-diffracting beam for fast imaging by projecting 3D volumes onto 2D images, but lacks depth details in these 2D images. (c) The MicroDiffusion model reconstructs 3D volumes from 2D projections obtained in volumetric imaging, enhancing imaging performance by combining rapid speeds with improved depth resolution.
\caption{Background and concept of MicroDiffusion-enabled volumetric microscopy. (a) Conventional 3D laser scanning microscopy, while depth-resolvable due to its point-scanning 3D data acquisition scheme, suffers from slow imaging speed. (b) Volumetric microscopy using a non-diffracting laser beam provides fast volumetric imaging by axially projecting 3D volumes onto 2D images but lacks depth information within each acquired 2D image. (c) Our proposed MicroDiffusion model is employed as a digital backend for 3D volumetric reconstruction from 2D projections acquired in (b). MicroDiffusion significantly enhances volumetric imaging performance, providing a synergistic balance between imaging speeds and depth-resolving capabilities.
% By using our proposed MicroDiffusion model as a digital backend for our acquired 2D images from volumetric imaging, we can reconstruct 3D volumes. This enhances volumetric imaging performance, offering both fast imaging speeds and depth-resolving capability.
}
\label{fig:definition}
\end{figure}
Volumetric optical imaging has emerged as a pivotal tool in biological and medical domains, enabling precise 3D visualization of intricate structures with unprecedented temporal resolution \cite{yun2006comprehensive, gottschalk2019rapid}. Despite its high spatial resolution, the predominant approach in optical microscopy, reliant on 3D laser scanning, suffers from suboptimal temporal resolution due to the slow data acquisition inherent in point-scanning methods (Fig.~\ref{fig:definition}(a)). This limitation not only restricts clinical diagnosis mostly to 2D imaging, potentially compromising diagnostic accuracy \cite{goedeke2019multiphoton, cho2011multiphoton}, but also impedes the observing of dynamic 3D biological processes \cite{huisken2004optical,helmchen2005deep}.

% Recent advances in optical imaging have sought to address this by introducing elongated nondiffractive laser beams \cite{durnin1987exact}, such as Bessel \cite{rodriguez2018three} and Airy beams \cite{tan2019volumetric}. Unlike conventional point scanning with Gaussian beams, these novel techniques offer significantly faster volumetric imaging rates \cite{rodriguez2018three, tan2019volumetric, cao2023optical}. Such advances have opened doors to observing events like neuronal firing across multiple axial planes in real-time, an enormous stride towards deciphering complex neuronal circuits \cite{lichtman2011big}. However, this innovation is not without challenges. 

Recent advancements using non-diffracting beams have expedited laser scanning microscopy by optically projecting 3D volumes as 2D projections for volumetric imaging~\cite{durnin1987exact, rodriguez2018three, tan2019volumetric}. However, this approach sacrifices depth information within each 2D snapshot~\cite{valle2019two, chen2018rapid} (see Fig.~\ref{fig:definition}(b)), necessitating the development of tools capable of reconstructing depth for accurate 3D reconstruction from 2D images. In this paper, we aim to reconstruct 3D volumes from limited 2D projections obtained through such volumetric imaging, striving to expedite optical volumetric imaging without compromising depth resolvability of 3D volumes.
% In this paper, our primary objective is to enhance the recovery of depth information from 2D projections by generating densely stacked slices along the axial axis to cover the entire 3D volume. 
% \yuyin{@Fei, check if my modification aligns with Figure 1 well}

% \yuyin{INR, though present a good global view/encodes global information, can result in noisy/incomplete . Diffusion, on the other hand, can better capture localized details UNet, xxxx. Theerefore, we propose INR as global prior, complemented with Diffusion to generate the finer details.}

% This resembles the computer vision task of 3D reconstruction and view synthesis from limited 2D views, which has seen significant progress with the seminal work on Neural Radiance Fields (NeRF)~\cite{mildenhall2021nerf}.
Existing 3D reconstruction methods, such as Implicit Neural Representations (INR)~\cite{sitzmann2020implicit}, offer a comprehensive global view from given 2D microscopy projections by mapping coordinates to a holistic 3D volume using neural networks. However, direct 3D reconstruction using INR often yields globally coherent yet visually blurry outputs, lacking local details. This limitation in spatial resolution may stem from the limited number of 2D images acquired.
Conversely, Denoising Diffusion Probabilistic Models (DDPM)~\cite{ho2020denoising}, especially with the U-Net architecture~\cite{ronneberger2015u}, excel in detailed generative modeling, managing spatial hierarchies, and preserving fine-grained details. 
Building upon these strengths, we propose MicroDiffusion, a hybrid approach integrating INR's global structural coherence with DDPM's detail enhancement capabilities. 

MicroDiffusion encompasses two key designs as shown in Fig.~\ref{fig:model}: 1) \textbf{INR pretraining}, which transforms 2D projections into a preliminary 3D volumetric output, establishing a global structure; and 2) \textbf{implicit representation-guided diffusion}, where the pretrained INR acts as a global prior guiding a diffusion model, enhancing details and reducing noise in local 2D projections within the 3D volume. 
Specifically, MicroDiffusion employs a linear interpolation of INR model output with the noise input, rather than starting from a conventional Gaussian noise baseline, to enrich the diffusion process with structured 3D information. 
Furthermore, this step enhances image fidelity by conditioning image and positional embeddings extracted from the closest projections. Thus, MicroDiffusion generates 3D reconstructions faithfully representing original optical microscopy images.

Comprehensive experiments on three optical microscopy datasets showcase MicroDiffusion's efficacy. 
Compared to the baseline INR, it notably enhances reconstruction quality by up to 15.5\% in SSIM, 15.2\% in PSNR, and 64.7\% in DICE on Dendrite dataset, up to 15.0\% in SSIM, 3.0\% in PSNR, and 0.3\% in DICE  on Vasculature dataset, and up to 1.8\% in SSIM, 0.8\% in PSNR, and 4.7\% in DICE  on Neuron dataset. The resulting 3D stacks demonstrate remarkable resolution, delineating individual dendrites (less than $1 {\mu}m$) and preserving coherent 3D structures—an achievement unattainable by the naive DDPM approach. These tangible outcomes establish MicroDiffusion as a pioneering framework for reconstructing high-quality 3D volumes from 2D projections in volumetric microscopy using non-diffracting beams, reconciling the trade-off between depth information and imaging speed.

\section{Related Works}
\paragraph{Laser scanning microscopy with non-diffracting beams for volumetric imaging.} Laser scanning microscopy has emerged as the gold standard in biomedical imaging. Commonly used in biomedical applications, imaging modalities such as multiphoton microscopy \cite{helmchen2005deep}, optical coherence microscopy \cite{lee2008bessel}, and photoacoustic microscopy \cite{cao2023optical} all share at least one laser scanning mode. However, a critical challenge in laser scanning microscopy lies in accelerating data acquisition without sacrificing resolution and depth, especially with the most widely used point-scanning methods that scan a tightly focused 3D laser point to collect volumetric data (see Fig.~\ref{fig:definition}(a)).

To address this, optical strategies using non-diffracting laser beams, notably Bessel and Airy beams, have been proposed. These beams generate an elongated and almost uniform axial point spread function. Scanning with non-diffracting beams essentially captures multiple axial layers in a single lateral scan, as opposed to the point-scanning method (see Fig.~\ref{fig:definition}(b)). For instance, they have been utilized for rapid volumetric imaging, enabling the real-time capture of dynamic biological processes \cite{cao2023optical, rodriguez2018three, chen2018rapid}. Despite their speed advantages, these strategies often compromise depth information, yielding 2D projections without detailed information on feature depths. To combine the speed benefits of non-diffracting beam scanning methods with the capability to discern depth information, we propose a deep learning model for this inference (see Fig.~\ref{fig:definition}(c)).

\paragraph{Implicit Neural Representations (INR).}
Implicit neural representations excel at modeling the forms of 3D objects, generating surfaces for 3D scenes, and capturing detailed 3D structures. Pioneering work such as GQN~\cite{GQN} utilizes a generative query network to learn scene representations from multiple perspectives. 
Building on this foundation, Mildenhall et al.~\cite{mildenhall2020nerf} introduce the seminal concept of Neural Radiance Fields (NeRF), which use a multi-layer perceptron to encode 3D scenes for view synthesis. Other works, such as Poly-INR~\cite{singh2023polynomial}, SIREN~\cite{seo2022siren} and LIIF~\cite{chen2021learning}, have employed periodic activation functions, significantly enhancing the quality and adaptability of image representations. 

Parallel to these advancements, the application of INR in medical imaging has shown remarkable potential. For example, ARSSR~\cite{ARSSR} and CoIL~\cite{sun2021coil} have adapted NeRF-like methods for super-resolution in medical images. NeRP~\cite{shen2023nerp} distinctively combines the inherent image information with the physics of sparse measurements to enhance medical image reconstruction. Cryodrgn~\cite{zhong2020reconstructing} and fpm-inr~\cite{zhou2023fpminr} are notable for reconstructing 3D volumes from 2D microscopy images. 
As a recent advancement, IDM~\cite{gao2023implicit} integrates INR and diffusion models by employing INR as the decoder of a diffusion model. In contrast, we leverage INR to generate continuous and interpretable 3D representations used as guidance for a diffusion model.
%, focusing on overfitting to specific spatial domains.

% \hongru{
% Our work is distinct from these studies as we have access to only a very sparse set of 2D projections, and unable to leverage pretraining or supervision from low- and high-resolution pairs.
% } 
% We incorporate INR to generate continuous and interpretable 3D representations for a fitted volume. Our training regimen involves emulating the data acquisition process by incorporating INR and constructing 2D projections from the fitted 3D volume, which serves as our unique form of supervision.
% However, these works are not like ours, as we are working on limited number of 2D projections, where the models can not see a ground-truth image during training. Thus, the pretraining or supervising low-resolution and high-resolution pairs strategies doesn't work in our problem settings. 
% \zihao{We employ INR for their capacity to generate continuous and interpretable 3D representations by overfitting to a specific spatial domain. Our training regimen involves emulating the data acquisition process by constructing 2D projections from our fitted 3D volume, which serves as our unique form of supervision. }

% We utilizing INR for its ability in generating continuous and interpretable 3D representation, which will further enhanced utilizing a diffusion pipeline.
\vspace{-3mm}
\paragraph{Diffusion Models.}
Diffusion models are currently at the forefront of generative model innovation. The Denoising Diffusion Probabilistic Model (DDPM)~\cite{DDPM} can incrementally convert Gaussian noise into coherent signals. Subsequent research has expanded on controlling the output of these models, primarily categorized into classifier-guidance~\cite{dhariwal2021diffusion} and classifier-free guidance~\cite{ho2022classifierfree, rombach2022high}. Recent studies demonstrate the versatility of diffusion models in creating content guidance from a variety of sources, including images, text, depth, video, and their combinations~\cite{brooks2023instructpix2pix, rombach2022high, girdhar2023imagebind, ramesh2021zeroshot, saharia2021image, kawar2023imagic}.

In 3D reconstruction, considerable efforts are made to produce 3D models from text prompts or 2D references~\cite{poole2022dreamfusion, karnewar2023holodiffusion, liu2023one2345, szymanowicz2023viewset, cheng2023sdfusion, xu2023dream3d, lin2023magic3d, shue20233d}. The approach most similar to ours is Magic123~\cite{qian2023magic123}, which utilizes a two-stage, coarse-to-fine framework to generate 3D models with reference images. MicroDiffusion differs from Magic123 in several aspects. First, Magic123 employs pretrained knowledge from models like Stable Diffusion~\cite{poole2022dreamfusion,rombach2022high} or Zero1-to-3~\cite{liu2023zero1to3} to generate reference views for training Neural Radiance Fields. In contrast, our MicroDiffusion has no such pretrained knowledge, and we must directly train the Implicit Neural Representation (INR) from projection. Second, diffusion in the Magic123's fine stage is applied solely to improve the mesh generated from NeRF, without considering the NeRF's information. In MicroDiffusion, our diffusion model focuses on learning 3D reconstructions, with the INR acting as a source of prior knowledge for global information and actively contributing throughout the training process.

% MicroDiffusion stands out from these approaches in two key ways. Firstly, within the domain of microscope imaging, we lack pre-trained general diffusion models equipped with prior knowledge, unlike those used for generating 3D models in the natural imaging domain\cite{poole2022dreamfusion, karnewar2023holodiffusion}. Secondly, our focus is on the diffusion model's denoising capability, which enables it to function as a decoder for Implicit Neural Representation (INR) models.

\section{Problem Formulation}

As depicted in Figure~\ref{fig:definition}(b), a non-diffracting beam creates a uniform point spread function along the axial direction with a limited width, offering an \( n \)-fold increase in imaging speed when the optical axial width of the non-diffracting beam is \( n \) times that of a conventional point-like axial profile width beam. However, this advantage in volumetric microscopy comes at the cost of depth information, as these beams optically project 3D volumetric information along the axial direction, leading to a lack of depth information in resulting 2D images. 
% Currently, there is no method available to extract depth information from these 2D images.

This study aims to develop a model $f$ capable of reconstructing a depth-resolved 3D volume from 2D projections {$\{\textbf{X}_i\}$}, obtained using non-diffracting beams. The objective is to achieve a reconstructed 3D volume $\mathbb{M}$ with image quality comparable to traditional point-scanning methods. The 3D stacks that can be acquired from point-scanning methods within the same $i$th sub-volume are represented as $\mathbb{M}_{i} = \{m_1^{i}, m_2^{i}, \ldots, m_n^{i}\}$. In non-diffracting volumetric imaging, 2D projections are axially downsampled by a factor of $n$ in $\mathbb{M}_{i}$, resulting in each projection $\textbf{X}_i$ being expressed as $\textbf{X}_i = \frac{1}{n}\sum_{k=1}^{n} m_k^i$. Hence, the problem is to find a model $f : \{\mathbf{X}_i\} \rightarrow \mathbb{M}$ to reconstruct depth-resolved 3D volumes from downsampled 2D projections.

\section{Method}
\label{sec:Method}
% One primary challenge encountered in this problem is the lack of a substantial dataset. This is due to the fact we are developing a new type of imaging machine, which makes it difficult to train a supervised super-resolution model capable of handling such tasks. Also as our ground truth is only given the mean of the high dense projections, traditional frame interpolation algorithm is still not helpful. Consequently, we turn our attention towards Neural Radiance Fields, where we hoping to represent the 3D field with a neural model.
In this section, we begin by revisiting key concepts of Implicit Neural Representations (INR) and present our INR design crafted for optical microscopy reconstruction in Sec.~\ref{sec:inr}. 
We then delve into MicroDiffusion, our implicit representation-guided diffusion model in Sec.~\ref{sec:ddpm}. 
% Subsequently, we provide a detailed explanation of the integration of diffusion models with INR. 
% We discuss the details of the inference stage in Sec.~\ref{sec:inference}.

\subsection{Implicit Neural Representation}
\label{sec:inr}

% \begin{figure*}[htbp]
% \centering
% \includegraphics[width=\linewidth]{fig/INR.pdf}
% \caption{The structure of the INR encoder. We first samples uniformly from the 3D coordinates. The 3D coordinates is encoded to higher frequency utilizing positional encoding and then the feature is mapped to the certain density at its position using a MLP. We then project the constructed 3D structure to compare with our 2D projection ground truth.}
% \label{fig:INR}
% \end{figure*}

\paragraph{Revisit INR for 3D Reconstruction.}

% In the domain of 3D reconstruction, Implicit Neural Representation (INR) methods have been developed to model continuous volume space. These methods utilize a function, typically a Multilayer Perceptron (MLP), denoted as \( f_{\text{inr}} \), to implicitly represent a 3D field. The function \( f_{\text{inr}} \) is defined over continuous 3D space and maps voxel coordinates to a predicted property, such as intensity or occupancy. The formulation is as follows:
INR methods utilize a function, typically a Multilayer Perceptron (MLP) denoted as \( f_{\text{inr}} \), to implicitly represent a 3D field. \( f_{\text{inr}} \) operates over continuous 3D space and maps coordinates to a predicted property, like intensity or occupancy, formulated as 
\begin{equation}
m_{inr} = f_{\text{inr}}(p(z)),
\end{equation}
where \({z} \) denotes normalized 3D coordinate within the range \([-1, 1]\) to ensure uniformity across the input space. \( p(\cdot) \) denotes the positional encoding that transforms 3D coordinates into a higher-dimensional space, 
\begin{figure*}[htbp]
\centering
\includegraphics[width=\linewidth]{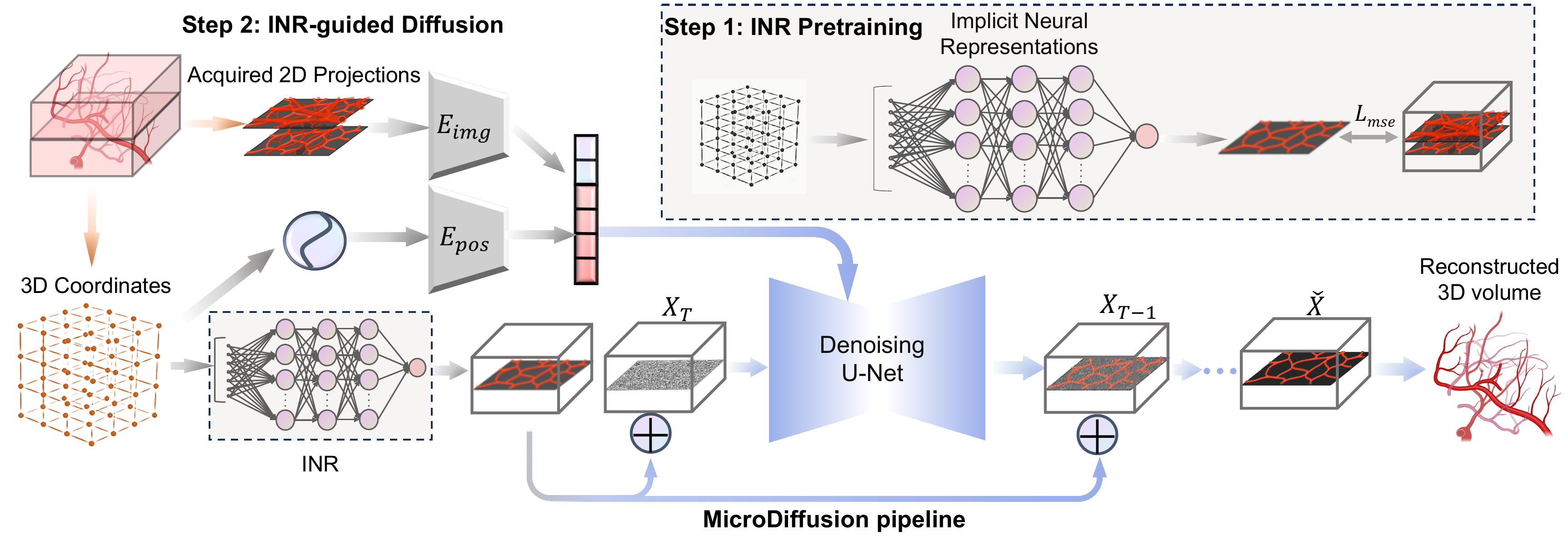}
\vspace{-0.3in}
\caption{Pipeline of MicroDiffusion. Step 1, we pre-train an INR which provides rough reconstructed images. Step 2, the 2D projections and 3D coordinates are used as the classifier-free guidance of the MicroDiffusion, and the INR output is integrated into the noisy image as guidance during the diffusion process. Detailed information is available at  Sec.~\ref{sec:Method}.}
\label{fig:model}
\vspace{-6mm}
\end{figure*}
crucial for capturing high-frequency details during reconstruction. $m_{inr}$ represents the property such as the intensity at position $z$.

Training INR involves a reference dataset, encompassing a set of 2D projections from 3D sub-volumes, as described in Sec.~\ref{subsec:dataset}, with 3D coordinates and corresponding intensities. The training objective is to minimize the reconstruction error between the predicted intensities \( m_{inr} \) and the actual data sampled at each coordinate. 

% To train INR, a reference dataset \( \mathbf{D} = \{(\mathbf{x},d)_i\} \), which is sparsely sampled representations of the 3D field, is employed. The training objective is to minimize the reconstruction error, defined as the difference between the predicted intensities \( \hat{d} \) and the actual sampled data \( d \) with the given coordinate.a

\paragraph{INR for Volumetric Microscopy Reconstruction.}

As shown in Figure~\ref{fig:model} (step~1), we sample 3D coordinates uniformly from the 3D volume $\mathbb{M}$, followed by positional encoding and the use of an MLP to map these encodings to voxel density values. For each reference projection $\mathbf{X}_i$, we compute the coordinates for \(n\) neighboring slices (defined as 2D projections of neighboring 3D sub-volumes), and concurrently synthesize these slices $\{m^i_1,\dots, m^i_n\}$ using \(f_{\text{inr}}\). 
Reconstruction loss is measured as the mean squared error (MSE) between the mean of the synthesized slices and the reference projection:
% In alignment with our specific problem setting, we have adapted the INR framework to suit our requirements as demonstrated in Fig. \ref{fig:INR}. Our approach simulates the scanning process by incorporating a projection step that aligns the distribution of our INR output with the acquired projection projections. For each reference projection \(X_i\), we compute the coordinates for \(N\) neighboring slices \(c_k\) and concurrently synthesize these slices \(m_k\) using \(f_{\text{inr}}\). The mean of these synthesized slices is then calculated and compared to the corresponding projection projections.
% The optimization process is guided by a loss function based on the Mean Squared Error (MSE) between the mean of the synthesized slices and the projections, which is formulated as:
\begin{equation}
L_{\text{mse}} = \sum_{i=1}^{N} \text{MSE}\left(\frac{1}{n} \sum_{k=1 }^{n} m^i_k, \mathbf{X}_i\right),
\end{equation}
where \(m_k\) denotes the \(k\)-th synthesized slice, and \(\mathbf{X}_i\) represents the \(i\)-th reference projection.

% We observe that the reconstructed slices for the 3D field look blurry with missing details. Furthermore, INR reconstructions contain some artifact parts which, as we conjecture, is rooted from the ambiguity of projection process.

\paragraph{INR Neighbouring-based Inference.}
\label{pg: Neighbour}
% In the training process, we simulate the projection process while calculating the optimization target loss. Thus, there could be a potential distribution shift when we predict the density of a certain position from solely its own coordinate. To tackle this issue, during inference, we align its distribution by using the neighboring slices of our target slice as references. For a particular synthesized slice denoted as \(x\), we estimate \(M\) neighboring slices, similar to the training process. Consequently, the inference for \(x_i\) is formulated as:

During training, we simulate the downsampling process triggered by the optical axial projection with a non-diffracting beam, and optimizate the target loss by averaging the output over $n$ coordinates. However, this may introduce a distribution shift when predicting the density solely from its own coordinate. To mitigate this issue, we incorporate information from neighboring slices during inference. When considering a particular 3D coordinate $z$, the inference result is obtained through a weighted average of $n$ neighboring slices. The weight for the $k$-th neighboring slice follows a Gaussian distribution:
\begin{equation}
g_k = \frac{1}{\sqrt{2\pi}} e^{-\frac{ (\frac{k}{n}-0.5)^2}{2}}
\end{equation}
% In the training process, we simulate the projection process while calculating the optimization target loss. Thus, there could be a potential distribution shift when we predict the density of a certain position solely from its own coordinate. To tackle this issue, during inference, we also utilize the neighboring slices. For a particular synthesized slice $m_z \in x$ sharing coordinate $\mathbf{z}$, we estimate \(n\) neighboring slices with the centering slice's axial coordinate as $\mathbf{z}$, similar to the training process. Consequently, the inference for \(m_z\) is
% \begin{equation}
% m_z = \sum_{k=1}^n g_k \hat{m}_k
% \end{equation}
% Here, rather than using a uniform mean, we apply a weighted mean, where \(g_k\) is the coefficient for each generated slice \(\hat{m}_k\). Slices \(\hat{m}_k\) that are closer to our target slice are assigned a higher weight. Typically, the coefficient follows a district Gaussian distribution:
% \begin{equation}
% g_k = \frac{1}{\sqrt{2\pi}} e^{-\frac{ (\frac{k}{n}-0.5)^2}{2}}
% \end{equation}
% \hongru{[hongru: can we be more specific on the weight coefficient, how are they determined?]}

While INR reconstruction offers a comprehensive global view, the reconstructed 3D slices suffer from blurriness, artifacts, and lack of fine details (as demonstrated in Figure~\ref{fig:sampling_slices}). These issues compromise the spatial resolution and overall reliability of optical microscopy. To address these limitations and ultimately improve reconstruction quality, we introduce a novel approach where we leverage INR as a global prior to guide a diffusion model, enhancing the details and reducing noise in each local 2D slice.

\subsection{Implicit Representation-Guided Diffusion}
\label{sec:ddpm}

\paragraph{Diffusion Models with Classifier-free Guidance } 
We employ Diffusion Models~\cite{sohl2015deep,ho2020denoising} to reconstruct 3D volumes.
As a likelihood model, Diffusion Model can gradually recover the data from Gaussian noise. 
The forward diffusion process transforms an input $X_0$ to Gaussian noise $X_T \sim \mathcal{N}(0,1)$ by $T$ iterations, defined as:
\begin{equation}
 q(X_t | X_0) = \mathcal{N}(X_t | \sqrt{\bar{\alpha}_t} X_0, (1 - \bar{\alpha}_t) I),
\label{eq:diff1}
\end{equation}
where $X_t$ represents the data with added noise at time step $t$. $\bar{\alpha}_t = \prod_{s=0}^{t} (1~-~\beta_s)$ and $\beta_s$ represents the noise variance schedule, and $\mathcal{N}$ represents the Gaussian distribution. 
During training, the neural network $\epsilon_\theta({X}_t, t)$ is trained to reconstruct the original data ${X}_0$ from the noised data ${X}_t$. This is achieved by minimizing $\ell_2$ loss between the predicted noise and the actual noise introduced in the data: 
\begin{equation}
L(\theta) = \mathbb{E}_{(X,t)} \left[ \| \epsilon - \epsilon_{\theta}(X_t, t) \|^2 \right]
\label{eq:loss1}
\end{equation}
where $t$ is time step in the forward diffusion process.
During the generation process, the neural network $\epsilon_{\theta}({X}_t, t)$ iteratively denoises ${X}_t$ to achieve high-quality output, $\theta$ is the trainable parameters of the model.

Classifier-free Guidance \cite{dhariwal2021diffusion,ho2022classifierfree} is a method for steering the output generation in Diffusion Models. Diverging from the standard approach of diffusion models, this technique involves training a neural network  $\epsilon_\theta({X}_t, t,c)$ with an additional conditioning $c$. The goal is to reconstruct $X_0$ while incorporating a probability $p_{\text{uncond}}$ that $c$~$\leftarrow$~$\emptyset$. The loss function $L(\theta)$ can be written as:
\begin{equation}
L(\theta) = \mathbb{E}_{(X,t)} \left[ \| \epsilon - \epsilon_{\theta}(X_t, t, c) \|^2 \right],
\label{eq:loss2}
\end{equation}
where $\emptyset$ is the null class. At the generation time, the model uses a guidance scale $\omega$ to balance the influence of the conditioning information. This is done by interpolating between the model's predictions with and without the conditioning:
\vspace{-0.2in}

\begin{equation}
\tilde \epsilon_{t} = \epsilon_\theta(X_t, t, c) + \omega \cdot (\epsilon_\theta(X_t, t, c) - \epsilon_\theta(X_t, t)),
\end{equation}
where $\tilde \epsilon_{t}$ is the noise distribution that model predicts.

%We first adopt this framework without any modification, where we take the projection's expected $\mathbf{z}$ coordinate as guidance and itself as the denoising target $X_0$.  Reconstructed slices manifest good quality of details. However, the training process is not stable, also the generated slices have no continuity in the 3D spatial domain.

% \hongru{[hongru: can we have this section to elaborate the basics for DDPM method, since it's our base model we should explain how it works. Then at the end of this part we can go with what you wrote ``...a naive diffusion model is almost incapable of learning any spatial continuity...'']}

% Through pilot study,we found that with only a limited number of 2D projections, a naive diffusion model is almost incapable of learning any spatial continuity and can only generate images that appear real but have no coherence in the 3D space. Therefore, we utilize INR's reconstruction as a prior knowledge which has encoded spatial relations.

\paragraph{Projection and Coordinate Guidance} 
\label{par:snap}
In MicroDiffusion, we use 2D projections and 3D coordinates as conditioning information $c$ in Eq.~\ref{eq:loss2}. Projections provide content information of the 3D volume while 3D coordinates provide 3D spatial information.
%INR prior primarily helps with global information and 3D coherence, however, for more detailed reconstructions, we need additional guidance which can produce more concise local-wise reference. 
%Therefore, we take the projection and spatial coordinates as conditioning information.
% \hongru{[hongru: could we add a short / one-sentence motivation for this modification? why we think they can help]}
% Besides the global information given by the INR prior, we would also take the projection and coordinates into consideration. As illustrated in Fig. \ref{fig:diff}, our guidance extractor has two separate encoders, an image encoder $E_{img}$ and a positional Encoder $E_{pos}$.  The $E_{img}$ takes the closest 2D projection projections as input, $E_{pos}$ take the positional encoding of $z$ as the input. The final condition embedding $c$ can be formulated as:

As depicted in Figure~\ref{fig:model}, we introduce two distinct encoders: an image encoder, denoted as $E_{img}$, and a positional encoder, denoted as $E_{pos}$. 
$E_{img}$ encodes the current projection $\mathbf{X}_z$, while $E_{pos}$ encodes the 3D coordinate $z$ of the current projection. The ultimate condition information $c$ is formulated as follows:
\vspace{-0.1in}

\begin{equation}
    c = E_{img}(\mathbf{X}_z) \oplus E_{pos}(p(z)),
\end{equation}
where $\mathbf{X}_z$ is the reference projection, $\oplus$ is the concatenate function and $p(\cdot)$ is the coordinate embedding function.

% The conditional embedding $c$ is then used as condition following classifier-free guidance diffuison model. The wrapped up training algorithm can be formulated as algorithm \ref{alg:diff_train} and the generation process can be formulated as $\ref{alg:diff_inference}$. We follows the symbols used in classifier-free guidance paper\cite{ho2022classifierfree}. Here $z_\lambda$ is the image interpolation at some time step $t$, $\epsilon_\theta$ is our diffusion model, $\emptyset$ is the null class.

\paragraph{INR Prior Integration} 
To leverage global information and coherent 3D structures in INR to guide the Diffusion Models, we integrate INR outputs as prior knowledge for the diffusion process. 
Specifically, the INR output $m_{inr}$ is linearly interpolated with the noisy image $X_t$, which is later to be denoised by Diffusion Models. 
In MicroDiffusion's training and testing process, for each noisy image $X_t$ at time step $t$, we perform linear interpolation with the INR output $m_{inr}$ pixel by pixel as
\vspace{-0.1in}
\begin{equation}
X_t' = \gamma m_{\text{inr}} + (1 - \gamma) X_t 
\end{equation}
where $X_t'$ is the INR-enhanced image, $X_t$ is the noised image that needs to be denoised by the diffusion model, $m_{\text{inr}}$ is the reconstructed output from INR, and $\gamma$ is the interpolation rate. This approach empowers the diffusion model to directly leverage structural information learned by INR, addressing the learning challenge with a limited number of input 2D projections and enhancing its capacity to generate images with correct 3D structures.

\paragraph{Training and Generation Process}
% \subsection{MicroDiffusion} 
MicroDiffusion adopts a conditional U-net~\cite{ronneberger2015u} similar to that in stable-diffusion~\cite{rombach2022high}. However, in our Denoising U-Net, we remove the cross-attention mechanisms, and add both time condition and conditional feature $c$ at each output of the ResNet block.
\begin{algorithm}
\caption{Training function of MicroDiffusion}
\label{alg:diff_train}
\begin{algorithmic}
\REQUIRE  $\mathbf{X}$: 2D projections; $z$: 3D Coordinate; $t$: Time step; $p_{\text{uncond}}$ : Probability of being unconditional.
    \STATE $m_{inr} = f_{inr}(p(z))$ : INR Inference in Sec.~\ref{pg: Neighbour}
    \STATE $c= E_{img}(\mathbf{X}) \oplus E_{pos}(p(z))$
    \STATE $c$ $\leftarrow$ $\emptyset$ with probability $p_{\text{uncond}}$
    \STATE $X_t \gets $sample from $q\left(X_t \mid X_{0}\right)$
    \STATE $X_t' = \gamma m_{inr} + (1 - \gamma) X_t$
    \STATE $L(\theta) = \mathbb{E}_{(X,t)} \left[ \| X_0 - \epsilon_{\theta}(X_t', t, c) \|^2 \right]$
    \STATE Take gradient step on $L(\theta)$
\end{algorithmic}
\end{algorithm}
MicroDiffusion training algorithm is formulated in Algorithm~\ref{alg:diff_train}, which continues running until convergence. $\oplus$ is the concatenate operation. During training, we first encode 3D coordinates and 2D projections into conditional features \(c\). We then generate the INR prior and the noised data \(X_t\), and linearly interpolate them with an interpolation rate \(\gamma\). After preparing all model inputs, we follow the equation \eqref{eq:loss2} to update the model parameters.

\begin{algorithm}
\caption{Sampling function of MicroDiffusion}
\label{alg:diff_inference}
\begin{algorithmic}
\REQUIRE $w$: guidance strength;  $z$: 3D Coordinate; $\gamma$: interpolation rate; $T$: Max time step; $\mathbf{X}$: 2D projections.
\vspace{-1em}
\STATE $X_T \sim \mathcal{N}(0,1)$
\STATE $c= E_{img}(\mathbf{X}) \oplus E_{pos}(p(z))$
\STATE $m_{inr} = f_{inr}(p(z))$ : INR Inference in Sec.~\ref{pg: Neighbour}
\FOR{$t=T$ to $1$}
\STATE $X_t' = \gamma m_{inr} + (1 - \gamma) X_t$
\STATE $\tilde \epsilon_t = (1+w)\epsilon_\theta(X_t',t,c) - w\epsilon_\theta(X_t',t)$
\STATE $ \epsilon_t = $ sample from $\tilde \epsilon_t$
\STATE $\tilde X_{t-1} = X_t -\epsilon_t $
\ENDFOR
\RETURN $X_0$
\end{algorithmic}
\end{algorithm}

Generation process is outlined in Algorithm~$\ref{alg:diff_inference}$. Here
$w$ is the condition weight controlling whether the model bias more towards conditional or unconditional generation. Similar to the training process, we first prepare all model inputs, and then have the model predict the noise distribution $\tilde \epsilon_t$ at the current time-step $t$. We sample a noise $ \epsilon_t$ from the noise distribution, subtract it from $X_t$, and repeat this for $T$ times. We repeat this process for all the coordinates until the algorithm converges.

% TODO:
% To make the generated slices more coherent, we also try to introduce an optical flow loss.

\section{Experiments}

\subsection{Datasets}
\label{subsec:dataset}
We collected experimental data using a conventional multiphoton laser scanning microscope, which has been a gold standard imaging tool for modern biomedical study. This approach is known for creating a three-dimensional, point-like point spread function. By 3D scanning the tightly focused Gaussian beam, we generated high-quality 3D volume stacks. These stacks serve as ground truth datasets for our research problem. Our setup captures 3D volume stacks of various biological structures—such as dendrites, neurons, and vasculature—within the shallow layers of the mouse cortex in a living animal (Fig.~\ref{fig:definition}). These datasets allow us to test our model with varied 2D projected images from diverse 3D biological features of varying densities.

Subsequently, we generated three synthetic datasets. These datasets simulate the case of fast data acquisition using a non-diffracting beam, whose point spread function has a quasi-uniform distribution axially and a predefined width (Fig. 1b). In later experiments, as we will demonstrate, we varied this width to be different multiples (denoted as step length 
$n$) of the Gaussian point spread function's axial width used to scan and generate the ground truth datasets. Consequently, the acquired 2D image sequences effectively averaged every $n$ frames along the axial direction, with no spatial overlapping in between. This approach reduced the volume data acquisition time by a factor of $n$. We then used both the ground truth and the generated synthetic datasets to evaluate our model at different step length $n$ values, focusing on various datasets from the brain. The design of these datasets allows us to directly determine the optimal step length $n$ value, which will inform both future optimal experimental data acquisition and hardware optical design.

\subsection{Implementation Details}
 Depending on the specific imaging modality, the practical axial width of a non-diffracting beam, such as a Bessel beam, can vary from a few times to tens or hundreds of times that of the point-like Gaussian beam used in conventional 3D laser scanning microscopes \cite{gottschalk2019rapid, chen2018rapid}. We initiate our experiments with a step length \( n \) of approximately 6, which corresponds to roughly an order of magnitude in speed-up — an important initial milestone for volumetric imaging. The performances of different reconstruction models were compared at this setting, and subsequently, an ablation study was conducted over the step length value. This study aims to further understand the impact of the step length of \( n \) on reconstruction quality, with the goal of identifying the optimal trade-off region between \( n \) times speed-up and image reconstruction quality.
  
 For computational efficiency, we downsample all the samples to a resolution of \( 128 \times 128 \) pixels in the lateral plane. For the pure Implicit Neural Representation (INR) model and the INR encoder, we map the 3D coordinates to a 512-dimensional space using a Gaussian-based embedding technique. The INR model is optimized using the Adam optimizer with a learning rate of \( 10^{-3} \) over 5000 epochs, a process that takes approximately 8 hours on an A-100 GPU. Additionally, we employ the AdamW optimizer with a learning rate of \( 2^{-4} \) and a weight decay of \( 10^{-4} \). As for the diffusion model, it is trained over 2000 epochs, taking around 4 hours on a single NVIDIA A-100 GPU.

\subsection{Baselines}
Given the novelty of this task and the absence of existing reference works, we established  baseline methods. The initial approach is a straightforward \textit{Interpolation} method, in which the generated structure is created through a uniformly weighted average of the two adjacent projections, weighted according to their distance. And \textit{Interpolation - cubic}, which estimates values by using cubic polynomials between points. This means that each interpolated curve segment is based on the position and slope (derivative) at its endpoints. The last baseline employs a pure INR method, which functions as our prior to the diffusion model.

\subsection{Reconstruction Results}

\subsubsection{Quantitative Results}             We evaluate our methods using three metrics: Peak Signal-to-Noise Ratio (PSNR), Structural Similarity Index Measure (SSIM), and the Dice coefficient. PSNR and SSIM are calculated slice-wise along the axial direction, with the mean value across all slices being reported. For the DICE coefficient, we  use the OTSU~\cite{otsu1979threshold} algorithm to determine the threshold for each image to assess the volumetric similarity between the generated 3D structure and the ground truth.
As presented in Table \ref{tab:main_result}, our method demonstrates strong performance, successfully capturing the principal structure of the original high-resolution model. Additionally, we observe that the diffusion model's decoder significantly enhances the performance of the pure INR model. %This improvement is particularly notable in the beads datasets, which are sparser compared to the other datasets.     

\begin{table}[h]
\centering
\footnotesize
\scalebox{0.85}{
\begin{tabular}{c|c|ccc}
   \toprule
   Dataset & Method & SSIM $\uparrow$ & PSNR $\uparrow$ & DICE $\uparrow$ \\
   \midrule
   \multirow{6}{*}{Dendrite} 
   &Interpolation         & 0.5799             & 28.78 & 0.6482 \\ 
   &Interpolation - cubic  & 0.6511 & 28.85 & 0.3973 \\ 
   &INR & 0.5837 & 25.81 & 0.4589   \\ 
   &Naive Diffusion  &  0.0297 & 19.95 & 0.2869 \\
    &Interpolation Diffusion&  0.6366 & 27.02 & 0.5729\\
    &Interpolation - cubic Diffusion& 0.4765 & 21.31 & 0.3786\\
   
   % &MicroDiffusion - join         & - & - & -  \\   
   &MicroDiffusion  & \textbf{0.6742} & \textbf{29.74} & \textbf{0.7557}  \\ 
   \midrule
    \multirow{6}{*}{Vasculature} 
   &Interpolation        & 0.3774             & 20.42 & 0.5936 \\ 
   &Interpolation - cubic  & 0.5204 & 20.52 & 0.4448 \\
   &INR                    & 0.5032 & 21.69 & 0.7136 \\  
   &Naive Diffusion  & 0.0207 &14.81 & 0.3234\\
    &Interpolation Diffusion& 0.4039 & 19.09 & 0.4860\\
    &Interpolation - cubic Diffusion& 0.2395 & 16.41 & 0.2672\\
   % &MicroDiffusion - join         & - & - & -  \\   
   &MicroDiffusion       & \textbf{0.5787} & \textbf{22.35} & \textbf{0.7158}  \\   
   \midrule
   \multirow{6}{*}{Neuron} 
   &Interpolation         & 0.1208  & 24.12 & 0.3553 \\ 
   &Interpolation - cubic  & 0.3265 & 26.50 & 0.1116 \\ 
   &INR                    & 0.4759 & 26.43 & 0.6403 \\
   &Naive Diffusion  & 0.0210 & 24.08 & 0.1468\\
    &Interpolation Diffusion& 0.4426 & 25.35 & 0.2425\\
    &Interpolation - cubic Diffusion& 0.3478 & 23.79 & 0.1318\\
   
   &MicroDiffusion       & \textbf{0.4845} & \textbf{26.66} & \textbf{0.6708} \\   
   \bottomrule
\end{tabular}}
\caption{Main results of the image reconstruction quality across different datasets with different biological features: vasculature, neurons, and dendrites. For all metrics, higher values indicating better performance as indicated by the arrows. }
\space{
}
\label{tab:main_result}
\end{table}
\vspace{-8mm}
\subsubsection{Qualitative Results}
\begin{figure}[htbp]
\centering
    \includegraphics[width=1\linewidth]{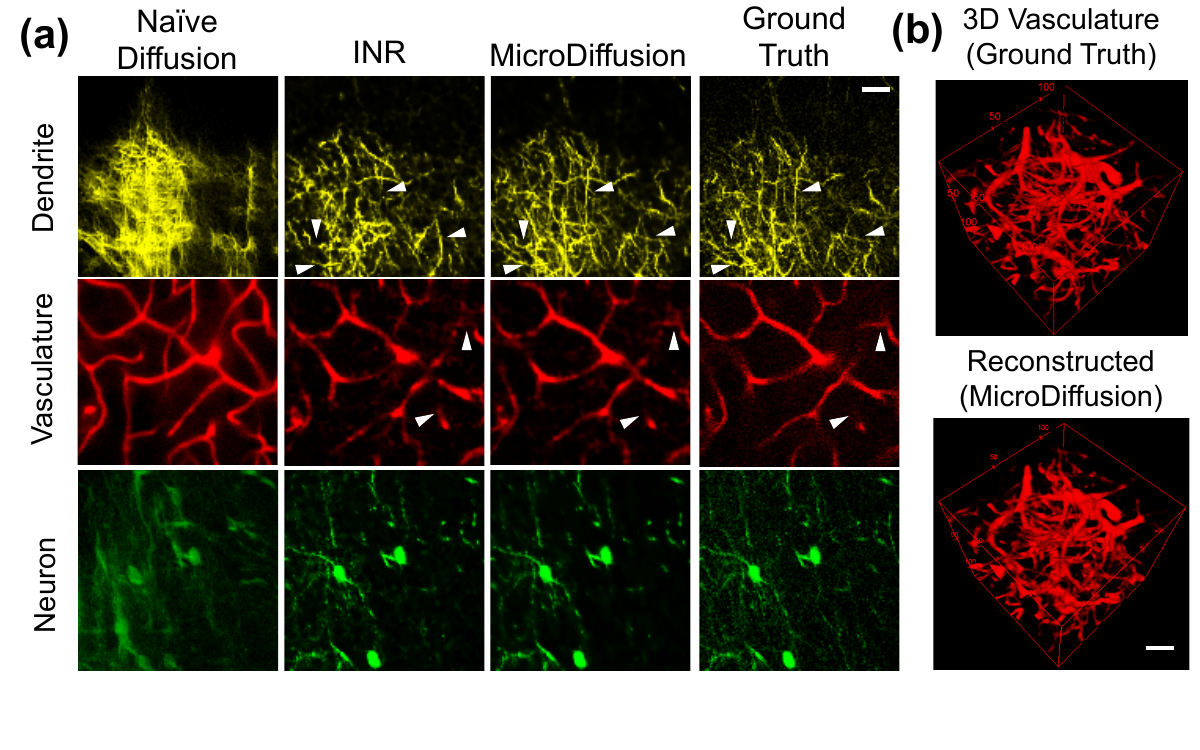}
    % \vspace{-6mm}
    \caption{Qualitative results: (a) Comparative visualization of slices from 3D reconstructions with different methods. Observable differences between the INR reconstruction, MicroDiffusion reconstruction, and ground truth are indicated with white arrows. (b) 3D vasculature. Scale bar: 30 $\mu$m.}
\label{fig:sampling_slices}
\vspace{-6mm}
\end{figure}

\vspace{-2mm}
Here, we present the reconstruction results of three methods: pure INR and MicroDiffusion. Part of the slices from the reconstructed 3D stacks are illustrated in Fig. \ref{fig:sampling_slices}, where we randomly selected three slices from the 3D reconstructions generated by naive diffusion, the pure INR reconstruction, our MicroDiffusion and compare with ground truth.  From these results, it is evident that the reconstructions obtained via the MicroDiffusion method more closely resemble the ground truth as the density of the biological features increases. This result indicates an encouraging possibility that volumetric imaging with a non-diffracting beam allows not only well-known volumetric imaging of sparse features such as neurons in the cortex \cite{chen2018rapid} but also denser features such as vasculature and even dense dendrites.

\subsection{Ablation study}
\subsubsection{Ablation on conditional feature}
\paragraph{How to encode 3D positional information?}
We initially evaluate two positional encoding methods for MicroDiffusion: (1) the sine-cosine based encoding as described by NeRF \cite{mildenhall2020nerf}, and (2) the Gaussian-based encoding used in NeRP \cite{shen2023nerp}. As shown in Table \ref{tab:position_ablation}, our experiment results demonstrate that the Gaussian-based encoding yields superior results, particularly in rendering clearer textures. We attribute this improvement to the intrinsic properties of Gaussian embeddings, which affords a more flexible mapping of positions to a higher-dimensional space. This flexibility enhances the subsequent learning process within MicroDiffusion, leading to more detailed and accurate representations.
\begin{table}[htbp]
\vspace{-3mm}
    \centering
    \resizebox{0.8\linewidth}{!}{
    \begin{tabular}{c|ccc}
    \toprule
        Method & SSIM$\uparrow$ & PSNR$\uparrow$ & DICE$\uparrow$ \\ 
    \midrule
        Sin-cos &  0.5243& 21.03& 0.6667\\ 
        gassian-based(ours) & \textbf{0.5787} & \textbf{22.35} & \textbf{0.7158} \\  
    \bottomrule
    \end{tabular}
    }
    \caption{Ablation of the positional encoding type on vasculature. }
    \label{tab:position_ablation}
\end{table}
\vspace{-7mm}
\paragraph{How to add conditional feature?}
We ablate on the way we incorporate the conditional feature $c$ into the model. \textit{w/o feature guidance} involves setting all conditions to $\emptyset$. \textit{cross-attention} involves adding $c$ using cross-attention to replace the self-attention in Denoising U-net, where the image feature is the query, and the conditional feature $c$ serves as the key and value. Experimental results demonstrate that adding the conditional feature $c$ is effective and leads to the best performance. This is likely because we have only one conditional feature, and in such a case, the cross-attention mechanism may not be effective. Therefore, it is better to directly add the conditional feature to the output of each ResNet block in the Denoising U-net.
\vspace{-3mm}
\begin{table}[htbp]
    \centering
    \resizebox{0.8\linewidth}{!}{
    \begin{tabular}{c|cccc}
    \toprule
         method & SSIM$\uparrow$ & PSNR$\uparrow$ & DICE$\uparrow$ \\ 
        \midrule
        w/o feature guidance & 0.5122 & 21.27 & 0.6512  \\ 
        cross-attention & 0.5371 & 21.25 & 0.6784 \\
         addition (ours) & \textbf{0.5787} & \textbf{22.35} & \textbf{0.7158} \\ 
    \bottomrule
    \end{tabular}
    }
    \caption{Ablation results fusion ablation on vasculature}
    \label{tab:fusion_ablation}
\end{table}
\vspace{-1mm}

\subsubsection{Ablation on INR prior}
\paragraph{How to generate INR prior?}
Here, we test three INR prior generation methods as shown in Table \ref{tab:neighbour_ablation}. \textit{no-neighbouring} means that we only utilize the INR output corresponding to the current 3D coordinate $z$ as the INR prior.
\textit{uniformly-mean} means that we use the uniformly averaged output of the INR corresponding to the six frames centered on the current 3D coordinate $z$.
The experimental results demonstrate that our approach performs the best when introducing Neighbouring-based Inference, allowing the model to obtain a more comprehensive 3D INR prior.

\begin{table}[htbp]
    \centering
    \resizebox{0.85\linewidth}{!}{
    \begin{tabular}{c|ccc}
    \toprule
        method & SSIM$\uparrow$ & PSNR$\uparrow$ & DICE$\uparrow$ \\ 
    \midrule
        no-neighbouring  & 0.4315& 15.26& 0.2025\\   
        uniformly-mean  & 0.4996 & 17.36 & 0.4086 \\ 
        Neighbouring-based (ours)  & \textbf{0.5787} & \textbf{22.35} & \textbf{0.7158} \\ 
    \bottomrule
    \end{tabular}
    }
    \caption{Ablation of neighbouring based inference on vasculature.}
    \label{tab:neighbour_ablation}
\end{table}
\vspace{-3mm}

\paragraph{How to add INR prior?} 
We investigate the necessity of the INR prior in this experiment. We trained a naive diffusion model that incorporates the INR prior as the projection introduced in~\ref{par:snap}.
We use the image encoder $E_{img}$ to encode the output of INR $m_{inr}$ and concatenate the feature with the other conditions.
This allows the model to generate images that resemble true biological features. However, this method performs poorly in acquiring global information, as evidenced by the very low DICE result in Table \ref{tab:prior_ablation}. 

% \vspace{-2mm}
\begin{table}[htbp]
    \centering
    \resizebox{0.75\linewidth}{!}{
    \begin{tabular}{l|ccc}
    \toprule
        Method & SSIM$\uparrow$ & PSNR$\uparrow$ & DICE$\uparrow$ \\ 
    \midrule
        Naive Diffusion &  0.4178& 14.97& 0.4540\\ 
        MicroDiffusion & \textbf{0.5787} & \textbf{22.35} & \textbf{0.7158} \\ 
    \bottomrule
    \end{tabular}
    }
    \caption{Ablation of diffusion model INR prior on the vasculature.}
    \label{tab:prior_ablation}
\end{table}

\vspace{-4mm}
\subsubsection{Ablation on training method}
\vspace{-1mm}
In our pipeline, we adopt a two-stage training process where the INR is trained initially and then frozen during the MicroDiffusion training. We conducted ablation experiments to explore two alternatives: (1) \textit{joint-training}, where we jointly train INR and the Denoising U-Net from random initialization, and (2) \textit{trainable}, where we unfreeze the INR during the MicroDiffusion training.

We used the same number of epochs for all methods. For \textit{joint-training}, we added the INR loss and applied a decaying weight to balance the training dynamics. As shown in Table \ref{tab:joint_ablation}, our method achieved the best performance. However, \textit{joint-training} proved to be too challenging and adversely affected the MicroDiffusion training process, while \textit{trainable} impaired the ability of INR to provide priors.

Therefore, we chose to train INR first and then train MicroDiffusion with it frozen.

 % \hongru{[What do we want to say here?]}
\begin{table}[htbp]
% \vspace{-2mm}
    \centering
        \resizebox{0.7\linewidth}{!}{
    \begin{tabular}{c|cccc}
    \toprule
         method & SSIM$\uparrow$ & PSNR$\uparrow$ & DICE$\uparrow$ \\ 
        \midrule
        joint-training & 0.5051 & 21.37 & 0.6603 \\
        trainable & 0.5347 & 21.27 & 0.6875  \\ 
        freeze (ours) & \textbf{0.5787} & \textbf{22.35} & \textbf{0.7158}  \\
    \bottomrule
    \end{tabular}
    }
    \caption{Ablation results of training methold on vasculature}
    \label{tab:joint_ablation}
\end{table}

\vspace{-8mm}

\subsubsection{Ablation on Different Step length}

We investigate the impact of the step length on MicroDiffusion performance. As outlined in our methodology, a larger step size results in faster volumetric imaging but makes reconstruction more challenging. Conversely, smaller step size leads to slower imaging but improved reconstruction. We keep the number of iterations the same and train our model from scratch. Results are presented in Figure \ref{fig:metrics}.

We observed that as the step length increased, all model metrics gradually decreased. To strike a reasonable balance between sampling speed and reconstruction quality, we chose a step length of 6.

% \begin{table}[htbp]
%     \centering
%     \begin{tabular}{cc|ccc}
%     \toprule
%         Step length $n$ & SSIM$\uparrow$ & PSNR$\uparrow$ & DICE$\uparrow$ \\ 
%     \midrule
%         4 &0.9894 & 24.31 & 0.8030 \\
%         6(ours) &0.9822 & 22.35 & 0.7158 \\  
%         8 &0.9801 & 21.91 & 0.6726\\
%         16 &0.9538 & 17.72 & 0.4259\\
%         32 &0.9266 & 15.70 & 0.2623\\
%         65 &0.9274 & 14.95 & 0.2030\\
%         130 &- & - & -\\
%     \bottomrule
%     \end{tabular}
%     \caption{Ablation of the step size on the vasculature. }
%     \label{tab: step_ablation}
% \end{table}

\begin{figure}[h]
\vspace{-3mm}
    \centering
    \begin{subfigure}[b]{0.45\textwidth}
        \includegraphics[width=\textwidth]{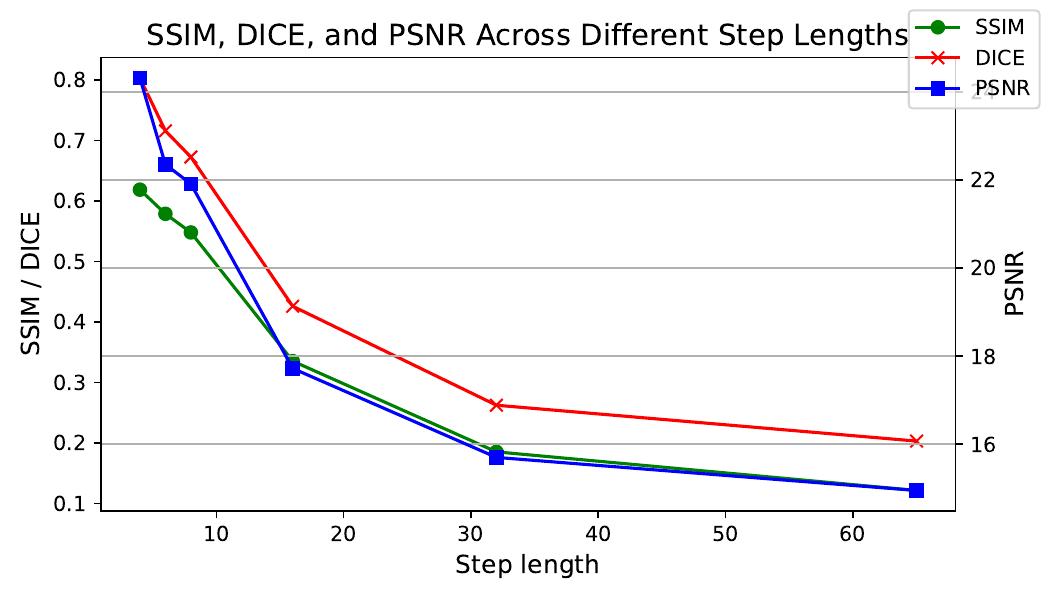}
        \caption{SSIM, PSNR and DICE}
        \label{fig:dice}
    \end{subfigure}
    \caption{Performance metrics across different step lengths.}
    \label{fig:metrics}
\end{figure}
 \vspace{-7mm}

\subsubsection{Reconstruction of sparse neuron dataset at various step lengths}
\begin{figure}
    \centering
    \includegraphics[width=1.0\linewidth]{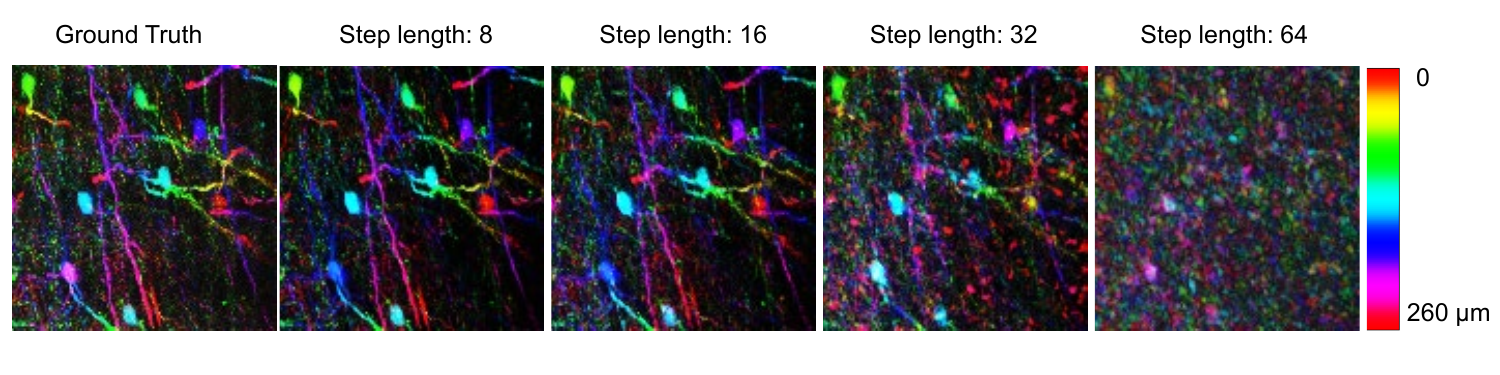}
    \caption{Reconstruction of the depth-resolved sparsely distributed neuron images and depth-resolved volumetric projections with different step lengths. }
    \label{fig:vis2}
        
\end{figure}

A natural question that may arise is whether we can further increase the step length if our features are sparse in space. To address this, Here, we conduct one further experiment aim to assess whether MicroDiffusion can further enhance the speed of volumetric imaging, particularly for samples with sparse spatial distribution. In this context, we have conducted a comparative analysis using the neuron dataset, which is the most sparse case among all the three datasets. The results of this comparison are illustrated in Figure~\ref{fig:vis2}. We evaluated the performance of our reconstruction models across a range of step lengths, which correspond to varying degrees of data acquisition speed. Notably, our findings indicate that, in the case of sparse neuron dataset, the step length can be extended to approximately 16 without significantly compromising the quality of the depth-resolved images. This suggests a potential for significant improvements in imaging efficiency without substantial loss in image fidelity for sparser featue of interest.
%\clearpage

\subsection{Discussion and Future Work}
In our experiments, we find that when ground truth data contains Gaussian noise, MicroDiffusion outperforms other methods in noise removal. This demonstrates the potential of MicroDiffusion for denoising 3D volumes acquired by volumetric optical microscopy. In cases where Gaussian noise is intentionally part of the Ground Truth data and should not be removed, it is necessary to investigate how to use other types of noise for MicroDiffusion training.

\section{Conclusion}
In this paper, we introduce MicroDiffusion, an innovative 3D reconstruction framework that adeptly addresses the challenges of rapid volumetric imaging and the need for depth-rich visualizations in biomedical research. By ingeniously integrating INR with DDPM, MicroDiffusion capitalizes on limited 2D projections to reconstruct high-resolution 3D images, significantly enhancing the capabilities of optical microscopy. Our approach not only accelerates the image acquisition process but also maintains 3D spatial information, allowing for the detailed observation of complex biological structures with minimal data acquisition at high speed. The successful application of MicroDiffusion across various datasets, from densely distributed dendrites to sparsely distributed neurons, underscores its potential as a transformative tool in medical diagnostics and fundamental biomedical research. This work paves the way for designing next-generation volumetric optical microscopy, setting a new benchmark for the integration of machine learning in 3D microscopy volume reconstruction, and opening avenues towards high-speed, high-resolution 3D optical microscopy.

\section*{Acknowledgement}
This work is partially supported by TPU Research Cloud (TRC) program, and Google Cloud Research Credits program.

{
    \small
    \bibliographystyle{ieeenat_fullname}
    \bibliography{main}
}

\clearpage
\section*{\Large{\textbf{Supplementary Material}}}
In this supplementary material, we detail the metrics we used in Sec.~\ref{sec:metrics}, including SSIM in Sec.~\ref{sec:SSIM}, PSNR in Sec.~\ref{sec:PSNR}, DICE in Sec.~\ref{sec:DICE}. Furthermore, we conduct an ablation study to find the best interpolation rate in Sec.~\ref{ab:rate}. Finally, we visualize the MicroDiffusion reconstruction results at various step lengths on the vasculature dataset in Sec.~\ref{vis:vas}.
\section{Details of the metrics}
\label{sec:metrics}

In this section, we delineate the metrics employed for assessing the quality of image reconstruction. We consider a reference image $x$ and a test image $y$, both being grey-level (8 bits) images of dimensions $M \times N$, drawn from respective sets $X$ and $Y$. Three evaluation measures are utilized: the Structural Similarity Index Measure (SSIM), the Peak Signal-to-Noise Ratio (PSNR), and the Sørensen–Dice coefficient (DICE), each detailed below.

\subsection{SSIM}
\label{sec:SSIM}
The Structural Similarity Index Measure (SSIM) serves as a pivotal metric in quantifying the resemblance between two images. It evaluates three fundamental aspects: Luminance, Contrast, and Structure.

\textbf{Luminance} is quantified through the mean gray scale value of the pixels, encapsulated in the equation:
\begin{equation}
l(x, y) = \frac{2\mu_x\mu_y + C_1}{\mu_x^2 + \mu_y^2 + C_1},
\end{equation}
where $\mu_x$ and $\mu_y$ represent the mean luminance of images $x$ and $y$, respectively. The constant $C_1$ prevents a zero denominator.

\textbf{Contrast} is gauged using the gray scale standard deviation, as:
\begin{equation}
c(x, y) = \frac{2\sigma_x\sigma_y + C_2}{\sigma_x^2 + \sigma_y^2 + C_2},
\end{equation}
where $\sigma_x$ and $\sigma_y$ denote the standard deviations of the images, and similarly, $C_2$ prevents a zero denominator.

\textbf{Structure} is assessed through correlation coefficients, formulated as:
\begin{equation}
s(x, y) = \frac{\sigma_{xy} + C_3}{\sigma_x\sigma_y + C_3},
\end{equation}
with $\sigma_{xy}$ being the covariance between $x$ and $y$. The constant $C_3$ ensures non-zero denominators.

The aggregate SSIM value, encapsulated within the range $[0,1]$, is derived as:
\begin{equation}
\text{SSIM}(x, y) = l(x, y) \cdot c(x, y) \cdot s(x, y),
\end{equation}
offering a comprehensive measure of similarity. Notably, a SSIM score of 0 implies an absence of correlation between the images, whereas a score of 1 indicates identical images. This index is particularly adept at capturing perceptual differences, making it a robust tool in image quality assessment.

\subsection{PSNR}
\label{sec:PSNR}
The Peak Signal-to-Noise Ratio (PSNR) is another crucial metric, predominantly focusing on the ratio between the maximum possible power of a signal and the power of corrupting noise. It is articulated as:
\begin{equation}
\text{PSNR}(x, y) = 10 \log_{10} \left( \frac{255^2}{\text{MSE}(x, y)} \right),
\end{equation}
where $\text{MSE}(x, y)$, the Mean Squared Error between the two images, is computed as:
\begin{equation}
\text{MSE}(x, y) = \frac{1}{MN} \sum_{i=1}^{M} \sum_{j=1}^{N} (x_{ij} - y_{ij})^2,
\end{equation}
with $x_{ij}$ and $y_{ij}$ representing the pixel values at the $ij^{th}$ position. The PSNR values range from 0 to $\infty$, where a higher value indicates superior image quality, reflective of lesser noise interference.

\subsection{DICE}
\label{sec:DICE}
The Sørensen–Dice coefficient (DICE), a statistical tool, quantifies the similarity between two sets. It is particularly effective in comparing the spatial arrangement of pixel values. The DICE is defined as:
\begin{equation}
\text{DICE}(X, Y) = \frac{2 |X \cap Y| + C}{|X| + |Y| + C},
\end{equation}
where $|X \cap Y|$ denotes the intersection size of sets X and Y, and $|X|$ and $|Y|$ are their respective sizes. The coefficient ranges from 0 to 1, with 1 indicating perfect agreement (complete overlap) and 0 denoting no overlap at all. This metric is particularly beneficial in scenarios where spatial correlation is a critical aspect of image similarity.

\section{What is the best linear interpolation rate?}
\label{ab:rate}
As demonstrated in the main paper, to incorporate global information and coherent 3D structures into the diffusion model, we employ a linear interpolation strategy between the Implicit Neural Representations (INR) output and the noisy image at each time step $t$. This approach is applied during both the training and testing phases of MicroDiffusion. Such integration of INR as prior knowledge is pivotal for guiding the diffusion process, particularly when dealing with limited 2D projection inputs.

An ablation study focusing on the interpolation rate $\gamma$ was conducted, with the results summarized in Figure~\ref{fig:metrics}. Our findings indicate that the effectiveness of $\gamma$ plateaus beyond a threshold of 0.1. Further increments in $\gamma$ yield minimal improvements, as evidenced by a marginal decrease in both Peak Signal-to-Noise Ratio (PSNR) and Structural Similarity Index Measure (SSIM) metrics. This trend suggests that, while the incorporation of INR prior is beneficial, an excessive reliance on it, particularly in the absence of Gaussian noise, can compromise the model’s generalization capabilities.

\begin{figure}[h]
\vspace{-3mm}
    \centering
    \begin{subfigure}[b]{0.45\textwidth}
        \includegraphics[width=\textwidth]{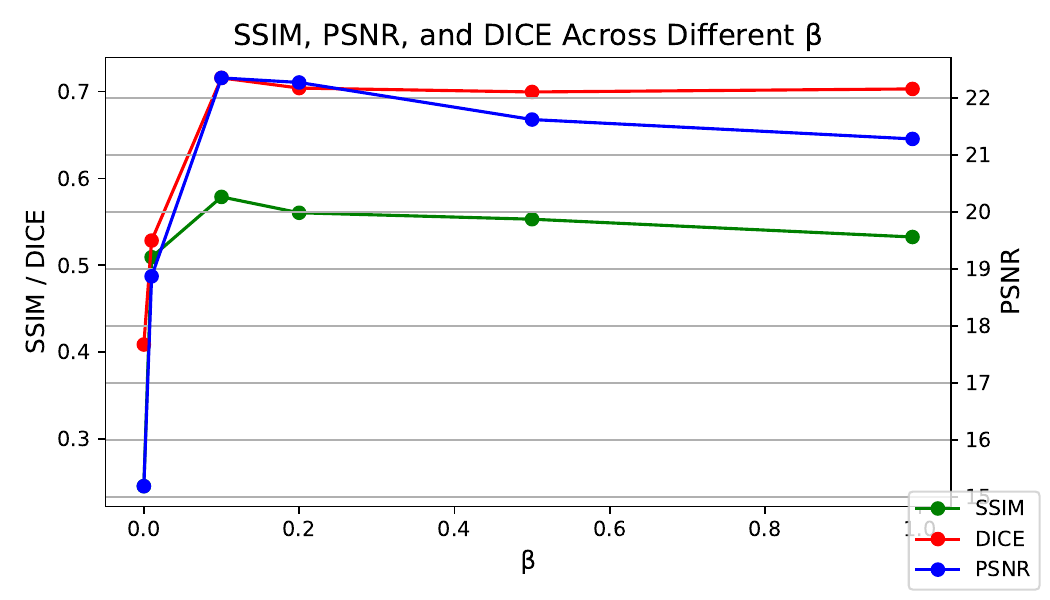}
        \caption{SSIM, PSNR and DICE}
        \label{fig:dice}
    \end{subfigure}
    \caption{Performance metrics across different linear interpolation rates.}
    \label{fig:metrics}
\end{figure}

\section{Visualization of MicroDiffusion reconstruction of vasculature at various Step Lengths}
\label{vis:vas}
\begin{figure}
\begin{minipage}{1\textwidth}
    \centering
    \includegraphics[width=1.0\linewidth]{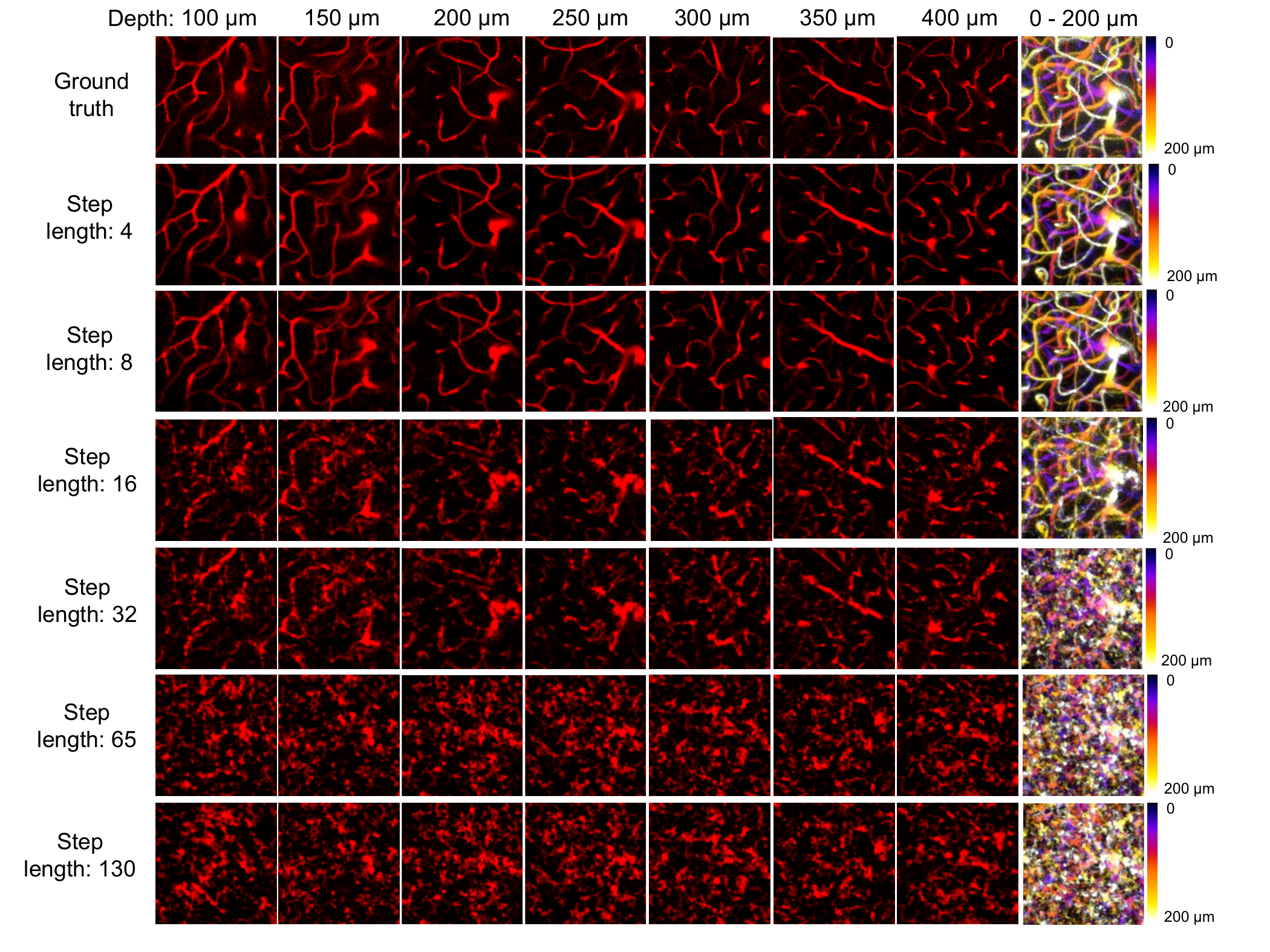}
    \caption{Reconstruction of the depth-resolved vasculature images and depth-resolved volumetric projections with different step lengths. }
    \label{fig:vis}
        \end{minipage}
\end{figure}

We visualize the reconstructed images in Figure~\ref{fig:vis}, which clearly demonstrates that the difficulty of reconstruction escalates with increasing step length, leading to a noticeable decline in model performance. This trend is quantitatively supported by the rapid decrease in metrics such as SSIM, PSNR, and DICE. Despite this challenge, it is noteworthy that satisfactory reconstruction quality is still achievable at step lengths of approximately 6 to 8. This finding is significant as it implies the potential to increase the speed of volumetric imaging by a factor of 6 to 8, enhancing imaging efficiency substantially. Looking ahead, our research aims to further improve model performance at even higher step lengths, pushing the boundaries of efficient and high-quality imaging in MicroDiffusion processes.

\clearpage

\end{document}